\documentstyle{article}

\begin{document} 
 
\title
{Cosmic ray diffusive acceleration at shock waves  with finite upstream 
and downstream escape boundaries} 
 
\author {M. Ostrowski$^{1,2}$ \& R. Schlickeiser$^2$ \\
$^1$Obserwatorium Astronomiczne, Uniwersytet Jagiello\'{n}ski, \\
 ul. Orla 171, 30-244 Krak\'{o}w,   
 Poland \\ (E-mail: mio{@}oa.uj.edu.pl) \\
$^2$Max-Planck-Institut f\"{u}r Radioastronomie, \,  Postfach 2024, \\ 
53010 Bonn 1, \, Germany \\ (E-mail: p337sch{@}mpifr-bonn.mpg.de) }
 
\date{}

\maketitle

\begin{abstract}
In the present paper we discuss the modifications introduced into the
first-order Fermi shock acceleration process due to a finite extent of
diffusive regions near the shock or due to boundary conditions leading
to an increased particle escape upstream and/or downstream the shock. In
the considered simple example of the planar shock wave we idealize the
escape phenomenon by imposing a particle escape boundary at some
distance from the shock. Presence of such a boundary (or boundaries)
leads to coupled steepening of the accelerated particle spectrum and
decreasing of the acceleration time scale. It allows for a
semi-quantitative evaluation and, in some specific cases, also for
modelling of the observed steep particle spectra as a result of the
first-order Fermi shock acceleration. We also note that the particles
close to the upper energy cut-off are younger than the estimate based on
the respective acceleration time scale. In Appendix A we present a new
time-dependent solution for infinite diffusive regions near the shock
allowing for different constant diffusion coefficients upstream and
downstream the shock. 
\end{abstract}

 
\section{Introduction} 
 
In the test particle approximation, the first-order Fermi shock
acceleration with infinitely extended diffusive regions upstream and
downstream the shock leads to a power-law particle spectrum with the
spectral index
 
$$\alpha = {3R \over R-1} \qquad , \eqno(1.1)$$ 
 
\noindent 
and the acceleration time scale 
 
$$T_{acc} = {3 \over U_1-U_2} \, \left( {\kappa_1 \over U_1} 
+ {\kappa_2 \over U_2} \right) \qquad , \eqno(1.2)$$ 
 
\noindent 
where the index "$1$" ("$2$") indicates respectively the upstream
(downstream) quantity, $U$ is the shock velocity, $R \equiv U_1/U_2$,
$\kappa$ is the spatial diffusion coefficient. For a review of the
results referring to the diffusive acceleration mechanism one should
consult some of the numerous review papers (e.g. Drury 1983; Blandford
\& Eichler 1987; Berezhko et al. 1988; Jones \& Ellison 1991).

In discussions of the acceleration process, the background conditions
are usually considered with particle diffusion coefficients changing at
most moderately with the distance from the shock. As a result, there are
infinite diffusive regions for cosmic ray particles. Then, particles are
removed from the acceleration region near the shock only due to
advection with the general plasma flow far downstream. However, if the
waves responsible for particle scattering are created due to the process
of cosmic ray streaming instability upstream the shock, e.g., the finite
amplitude waves resonant with particles near the spectrum cut-off energy
could be created only close to the shock within a finite time available
for the acceleration process. The analogous situation is the case if the
shock propagates through the finite volume of turbulent plasma. Then,
the energetic particles diffusing far from the shock will encounter the
conditions enabling them to permanently escape from the shock. A finite
extent of the shock wave to the sides and some particular boundary
conditions may also allow for such escape. The situation can be
qualitatively modelled by introducing the upstream (Berezhko et al.
1988) and/or downstream boundary for the energetic particle escape.
 
\begin{figure}
\vspace{6cm}
\caption{The configuration of the shock with the {\bf u}pstream {\bf f}ree
{\bf e}scape {\bf b}oundary ("u.f.e.b.") and the {\bf d}ownstream {\bf 
f}ree {\bf e}scape {\bf b}oundary ("d.f.e.b."). The situation is 
presented as seen in the shock rest frame. Between the shock and a given 
escape boundary there is a diffusive layer with some finite diffusion
coefficient $\kappa$. Behind the boundary one assumes infinite diffusion 
coefficient allowing for a free streaming to infinity (escape) of
particles hitting the boundary.} 
\end{figure} 
 
The conditions mentioned above are sometimes discovered in analysing
{\it in situ} measurements near heliospheric shock waves. Then, an
attempt to analyse the data within the standard approach, involving
equations (1.1-2) to describe the first-order Fermi acceleration, may
fail. For example, basing on such an analysis Bialk \& Dr\"oge (1993)
rejected the possibility of the first-order acceleration at the
considered shock wave and suggested the second-order acceleration
downstream the shock to play a role, instead. Unfortunately, to date
there is no theory available describing effects of the enhanced particle
escape at the acceleration process and discussions of the 'difficult'
cases have to be based on numerical methods. As a step forward, in the
present paper we develop a simple analytic theory allowing for
evaluation of measurements in terms of the diffusive length scales for
energetic particles. It can become a starting point for more elaborate
computations or numerical modelling. The theory describes modifications
introduced into the acceleration process due to finite extent of
diffusive regions near the shock (Figure~1). Below, in Section 2 we
discuss a simple time-dependent solution of the diffusion equation for
cosmic ray particles at the shock with the upstream and/or the
downstream escape boundary. To obtain analytic solutions, we consider a
simple case with constant diffusion coefficients, leading in the
stationary situation to power-law cosmic ray spectra. It enables us to
discuss the relation between the spectrum inclination and other
parameters of the acceleration process.  In Section 3,  we derive the
particle spectral index and the acceleration time scale as a function of
the boundary distance from the shock. The introduced particle sinks at
escape boundaries lead to steepening of the spectrum accompanied by a
substantial -- a factor of two or three for reasonable spectral indices
-- decrease of the acceleration time scale. We discuss the dependence of
these quantities on the upstream and downstream boundary distance from
the shock. Finally, in Section 4, we shortly summarize the results. The
presented theory allows one to interpret the above mentioned Bialk \&
Dr\"oge (1993) data in terms of the first-order acceleration. We also
note that the particles close to the upper spectrum energy cut-off are
younger than the estimate based on the respective acceleration time
scale valid at infinite times. The new time-dependent solution for the
particle distribution function in the case of infinite boundary distance
is presented in Appendix A. It is a generalization of the Toptygin
(1980) solution for the case of different constant diffusion
coefficients upstream and downstream the shock.
 
\section{On solution of the time-dependent problem} 
 
The problem of time-dependent solution of the diffusion equation for 
cosmic ray particles accelerated at the shock has been discussed by Drury 
(1983, 1991). Below, we will follow the approach described in the 
former of the mentioned papers. In order to produce power-law particle
momentum spectra with the finite extent of free escape boundaries one 
must restrict considerations to the momentum-independent diffusion 
coefficient. As in such conditions the spatial dependence of $\kappa$ 
does not lead to any qualitative changes of the acceleration process, we
consider the most simple situation with the diffusion coefficient 
spatially constant, separately upstream and downstream. As mentioned 
previously, we consider the particle acceleration process in the 
test-particle approximation at the non-modified shock structure, i.e. 
with flow velocities constant outside the shock. 
 
Let us consider a planar shock wave propagating along the $x$-axis of
the reference frame, with the velocity in the upstream plasma frame 
directed toward decreasing $x$ values. In the text below we will refer 
to the spatial co-ordinates in the shock rest frame, where the 
co-ordinate $x$ upstream the shock is negative. We are going to solve the
diffusion equation for the isotropic part of the cosmic ray distribution 
$f$ = $f(t,x,p)$: 
 
$${\partial f \over \partial t} + U_i {\partial f \over 
\partial x} - \kappa_i {\partial^2 f \over \partial x^2} 
= Q \qquad , \eqno(2.1)$$ 
 
\noindent 
where $i$ = $1$ or $2$ and the mono-energetic source function is taken 
as 
 
$$Q(t,x,p) = Q_0 \delta (x) \delta (p-p_0) H(t) \qquad , \eqno(2.2)$$
 
\noindent 
where $Q_0$ is a constant, $\delta$ is the Dirac delta function and $H$
is the Heaviside step function. We look for the solutions satisfying the 
upstream ($x = -L_1$) and downstream ($x = L_2$) free escape boundary 
conditions 
 
$$f(t,-L_1,p) = 0 = f(t,L_2,p) \qquad , \eqno(2.3)$$ 
 
\noindent 
and the matching conditions at the shock ($x = 0$): 
 
$$\left[ f \right] = 0 \qquad , \eqno(2.4)$$ 
 
$$\left[ \kappa {\partial f \over \partial x} + {U\over 3} p 
{\partial f \over \partial p} \right] = - Q_0 \delta (p-p_0) H(t) 
\quad , \eqno(2.5)$$ 
 
\noindent 
where the square brackets denote the differences between the upstream 
and downstream values of the given quantity across the shock. Following 
a standard approach described by Drury (1983) we start with the Laplace 
transform with respect to time of the distribution $f$, 
 
$$g(s,x,p) = \int_0^\infty e^{-st} f(t,x,p) dt \qquad , 
\eqno(2.6)$$ 
 
\noindent 
leading to the following form of Equation~(2.1) at $p > p_0$:
 
$$ s g + U_i {\partial g \over 
\partial x} - \kappa_i {\partial^2 g \over \partial x^2} 
= 0 \qquad . \eqno(2.7)$$ 
 
\noindent 
The solutions to this linear equation have the form $g(x)$ $=$ $C_+ 
\exp{(\beta_{i,+}x)}$ $+$ $C_- \exp{(\beta_{i,-} x)}$ ($i$ = $1$, $2$; 
$C_\pm$ = const). With the definition $\tau_i \equiv 4 \kappa_i/U_i^2$,
the exponents $\beta_i$ are given as 
 
$$\beta_{i,\pm} = {2\over \tau_i U_i} \left( 1 \mp 
\sqrt{1+\tau_i s} \right) \qquad . \eqno(2.8)$$ 
 
\noindent 
Boundary conditions (2.3) for the function $g(s,x,p)$ are
$g(s,-L_1,p) = 0 = g(s,L_2,p)$ and the upstream and downstream solutions 
of Equation~(2.7) can be written as
 
$$g_1(s,x,p) = C_1(s,p) \left[ e^{\beta_{1,-} x} - 
e^{-{4\sqrt{1+\tau_1s} \over \tau_1 u_1} L_1} 
e^{\beta_{1,+} x} \right] \quad , \quad\,\,\,\,\, \eqno(2.9)$$

$$g_2(s,x,p) = C_2(s,p) \left[ e^{\beta_{2,+} x} - 
e^{-{4\sqrt{1+\tau_2s} \over \tau_2 u_2} L_2} 
e^{\beta_{2,-} x} \right] \quad . \eqno(2.10)$$ 
 
\noindent 
One may note that, contrary to the solutions with no escape boundary,
the present ones make use of both the indices $\beta_+$ and $\beta_-$.
The matching conditions at the shock for the function $g(s,x,p)$ are
derived from Equations (2.4,5) as
 
$$\left[ g \right] = 0 \qquad , \eqno(2.11)$$ 
 
$$\left[ \kappa {\partial g \over \partial x} + {U\over 3} p 
{\partial g \over \partial p} \right] = - {1 \over s} 
Q_0 \delta (p-p_0) \quad , \eqno(2.12)$$ 
 
\noindent 
By imposing the condition (2.11) at the solutions (2.9,10) we obtain
 
$$g_1(s,x,p) = g_0(s,p) { e^{\beta_{1,-} x} - 
e^{-{4\sqrt{1+\tau_1s} \over \tau_1 u_1} L_1} 
e^{\beta_{1,+} x} \over 1 -e^{-{4\sqrt{1+\tau_1s} \over \tau_1 U_1}L_1} 
} \quad , \eqno(2.13)$$

$$g_2(s,x,p) = g_0(s,p) { e^{\beta_{2,+} x} - 
e^{-{4\sqrt{1+\tau_2s} \over \tau_2 u_2} L_2} 
e^{\beta_{2,-} x} \over 1 -e^{-{4\sqrt{1+\tau_2s} \over \tau_2 U_2}L_2} 
} \quad , \eqno(2.14)$$ 
 
\noindent 
where $g_0(s,p) \equiv g(s,0,p)$ is the Laplace transform of the 
distribution function at the shock. Its functional form can be derived 
with the use of the condition (2.12) and Equation~(2.8) as
 
$$g_0(s,p) = {3 Q_0 \over (U_1-U_2) s } 
\left( {p_0 \over p} \right)^{\alpha (s)} H(p-p_0) 
\qquad , \eqno(2.15)$$ 
 
\noindent 
where 
 
$$\alpha(s) = {3 R \over 2(R-1)} \left\{ 1+ \sqrt{1+\tau_1s} 
\, \coth{\left( {2L_1\sqrt{1+\tau_1s} \over \tau_1 U_1 } \right) } 
\right. \qquad$$ 
$$ \qquad \qquad \qquad \left. + \, {1 \over R}
\left[ \sqrt{1+\tau_2s} 
\, \coth{\left( {2L_2\sqrt{1+\tau_2s} \over \tau_2 U_2 } 
\right)} -1 \right]  \right\} \qquad . \eqno(2.16)$$ 
 
The distribution $f(t,x,p)$ can be formally derived by inverting the 
respective upstream and downstream transforms (2.13) or (2.14): 
 
$$f_j(t,x,p) = {1 \over 2\pi i} \int_{-i\infty}^{+i\infty} 
e^{st} g_j(s,x,p)\, ds \qquad (j = 1, 2) \quad , \eqno(2.17)$$ 
 
\noindent 
where the path of integration lies to the right of all singularities of
the integrand. The result of such inversion for a particular case of
$L_1 = L_2 = \infty$, and for constant diffusion coefficients which can
be chosen independently upstream and downstream and not necessarily be
equal, $\kappa_1 \ne \kappa_2$, we present in Appendix A. However, even
without actually performing the integration one is able to extract from
Equations~(2.13-15) quite a lot of information (Drury 1983). The
asymptotic behaviour at large times is obtained by looking just at the
contribution of the rightmost singularity of the integrand, here a
simple pole at $s = 0$. It gives the power-law steady state spectrum at
the shock, $f(\infty, 0, p) \equiv f_0(p)$, with the spectral index
$\alpha_0 \equiv \alpha(0)$:
 
$$\alpha_0 = {3R \over R-1} \left\{ {1 \over 
1-e^{-{U_1 L_1 \over \kappa_1}}} + { 1 \over R}\, 
{e^{-U_2 L_2 \over \kappa_2} \over 1 - 
e^{-U_2 L_2 \over \kappa_2} } \right\} \quad . \eqno(2.18)$$ 

\noindent 
The first term in the product at the right-hand side represents the 
spectral index $\alpha$ for the shock with infinite diffusive regions 
(cf. Equation~1.1), and the two parts of the second term describe
modifications due to particle escape through the upstream and downstream 
boundaries. 
 
With the use of Equations~(2.15,16,18), in the solution (2.17) taken at the
shock ($x \, = \, 0$) one can separate a part representing the limiting 
stationary solution $f_0(p)$ multiplied by a factor describing the time 
dependence of the full solution: 
 
$$f_0(t,p) = f_0(p) \cdot \int_0^t \psi (t^\prime) dt^\prime 
\qquad , \eqno(2.19)$$ 
 
\noindent 
where, with the notation $\Delta \alpha (s) \equiv \alpha(s) -\alpha_0$ ,
 
$$\psi (t) = {1 \over 2\pi i} \int_{-i\infty}^{+i\infty} 
e^{st} \left( {p_0 \over p} \right)^{\Delta \alpha (s)} \, 
ds \qquad . \eqno(2.20)$$ 
 
\noindent 
One can readily show that the integral
 
$$\int_0^\infty \psi (t) dt = 1 \qquad . \eqno(2.21)$$ 
 
\noindent 
If we denote 
 
$$h(s) = \exp{\left[ \Delta \alpha (s) \ln{(p/p_0)} \right]} 
\eqno(2.22)$$ 
 
\noindent 
the mean acceleration time from $p_0$ to $p$ is given as 
 
$$T_{acc}(p,p_0) \equiv \int_0^\infty t \psi (t) dt = 
{d \, h \over ds}(s=0) \qquad . \eqno(2.23)$$ 
 
\noindent 
With Equations~(2.16, 18.22) and assuming $\ln (p/p_0) = 1$ the above
formula yields the acceleration time scale
 
$$T_{acc} = {3R \over R-1} \, \left\{ 
{\kappa_1 \over U_1^2}\, 
\coth{ \left( {U_1 L_1 \over 2 \kappa_1} \right) } \, - \, {L_1 \over 
2U_1 \sinh^2{\left( {U_1 L_1 \over 2 \kappa_1} \right) }}  \,\, + 
\right. \qquad$$ 
$$\qquad \qquad \left. 
{\kappa_2 \over R U_2^2}\, 
\coth{ \left( {U_2 L_2 \over 2 \kappa_2} \right) } \, - \, {L_2 \over 
2RU_2 \sinh^2{ \left( {U_2 L_2 \over 2 \kappa_2} \right) } } 
\right\} \qquad . \eqno(2.24)$$ 
 
\noindent 
For $L_i \rightarrow \infty$ ($i$ = $1$, $2$) the expressions (2.18,24) 
reduce to the standard formulae (1.1,2).

\section{The acceleration process with free escape boundaries} 
 
Let us consider variations of the particle acceleration time scale (2.24) 
and the spectral index (2.18) with the escape boundary distance. For the 
discussion, we choose the simplest conditions with $\kappa_1 = 
\kappa_2$ and either $L_1 = L_2 = L$, or one $L_i = \infty$, but the 
generalization to more general conditions is a straightforward one. 
Below, a boundary distance is expressed in the units of the respective 
diffusive length scale, $L_{diff,i} = \kappa_i/U_i$ ($i$ = $1$, $2$). 
 
\begin{figure}
\vspace{6cm}
\caption{The dependence of the characteristic acceleration time, $T_{acc}$, 
normalized to the acceleration time scale at the shock with infinite 
diffusive regions, $T_{acc,0}$, versus the distance $L$ of the free escape 
boundary (given in units of the respective diffusive scale $\kappa_i / 
U_i$, $i$ = $1$ or $2$). Full lines show the relations for the case of 
equally distant upstream and downstream escape boundary ($L_1$ = $L_2$ = 
$L$) with different compressions: $R = 4$ (lower line), $R = 3$ 
(intermediate line) and $R = 2$ (upper line). The lines with long dashes 
represent the above relation for downstream particle escape ($L_1 = 
\infty$, $L_2 \equiv L$) and those with short dashes for upstream particle 
escape ($L_1 \equiv L$, $L_2 = \infty$). For picture 
clarity, in the last two cases we give curves for our limiting 
compressions, the lower curve for $R = 4$ and the upper curve for $R = 
2$.} 
\end{figure} 
 
\begin{figure}
\vspace{6cm}
\caption{The dependence of the stationary particle spectral index 
$\alpha_0$ versus the distance $L$ of the free escape boundary (given in 
units of the respective diffusive scale $\kappa_i / U_i$, $i$ = $1$ or 
$2$). The meaning of different curves is the same as at Figure~2. Here,
the shock compression values are given near the respective curves.}
\end{figure} 
 
\begin{figure}
\vspace{6cm}
\caption{The dependence of the characteristic acceleration time ratio 
$T_{acc}/T_{acc,0}$ (cf. Figure~2) versus the spectral index $\alpha_0$
(cf. Figure~3). The meaning of different curves is the same as at
Figure~2, the shock compression values are given near the respective curves. }
\end{figure}
 
In Figure~2, we present the acceleration time scale variations due to
changing the escape boundary distance.  A substantial decrease of this
time scale at small $L$ should be noted.  The reason for that is clear.
The particles which could diffuse for a long time in the case with
escape boundaries in infinity, for finite $L$ will probably disappear
from the acceleration process as their chance to wander far off the
shock is higher.  So, with small $L$, the spectrum is set up only by
particles diffusing very close to the shock with a short mean time
between the successive shock crossings, resulting in the short
acceleration time. The dashed lines show how the acceleration time is
affected by placing the escape boundary upstream only or downstream
only. Here and in the figures below, one may note an apparent asymmetric
influence of the upstream and the downstream boundary. The difference
results from the fact that the mean time between successive particle
interactions with the shock is related to the diffusive scale different
on both sides of the shock. In Figure~3, variations of the stationary
spectral index $\alpha_0$ are presented. The expected steepening of the
spectrum with decreasing $L$ due to the accompanied increased particle
escape is presented.  For larger $L$, the curves quickly converge to the
limiting values $3R/(R-1)$.  Imposing the escape boundary downstream
only has a much smaller influence on the spectral index than the
upstream boundary at the same distance (in our units of $L_{diff}$) from
the shock. This asymmetry can be easily understood by comparing the mean
times for particles at the shock to reach the escape boundaries. For
small $L_i$ ($i$ = $1$ or $2$), when the particle diffusive streaming
dominates over advection, this time scale can be estimated as $t_i
\approx L_i^2/\kappa_i$. For the assumed equal values for upstream and
downstream diffusion coefficients the respective diffusive distances
relate as $L_{diff,2} = R\, L_{diff,1}$ and $t_2 = R^2 t_1$ . So, the
escape through the upstream boundary will be a factor $t_2/t_1 = R^2$
greater than the downstream escape, and thus it will be mostly
responsible for increasing the spectrum inclination. For the same reason
the downstream boundary will influence the acceleration time scale to
the higher degree. For larger $L_i$ particle advection with the
general plasma flow  modify these estimates, but our conclusions are
still true. Always the boundary allowing for quick particle escape
preferentially acts to increase the spectrum inclination, while the
other one will provides the lower limit for the acceleration time scale.
 
A comparison of the spectrum and the acceleration time scale at various
$L$ is presented at Figure~4.  Along the curves, the parameter $L$
decreases to the right, i.e.  with increasing $\alpha$.  One may note
that steepest parts of the curves are to the left, near the limiting
spectral index. For all the considered shock compression values it is
possible to decrease $T_{acc}$ by about two times (more for larger $R$,
or if the particle escape boundary occurs only downstream) when the
spectral index steepens on $\Delta \alpha \approx 1.0$ .
 
\section{Discussion} 
 
The presented results allow one to model a wider range of cosmic ray
spectra as the output of the first-order Fermi acceleration at shock
waves, including distributions with a spectral index steeper than the
value given by Equation~(1.1). One should note that accelerating
particles with a steep spectrum due to finite escape boundaries allows
to obtain spectra extended to somewhat higher energies, due to decrease
of the acceleration time scale with respect to the one given in
Equation~(1.2). Another point to be mentioned here is an apparent
asymmetry of the acceleration process with respect to imposing the
escape boundary upstream or downstream the shock. We would like to note,
however, that the respective influence of these boundaries discussed in
the previous section can change if the downstream particle diffusion
coefficient is much smaller than the upstream one. In general, the
boundary allowing for quick particle escape will always preferentially
act to increase the particle spectrum inclination, while the other one
will limit in a higher degree the acceleration time scale. Let us also
note the fact that a downstream boundary further away than
approximately two diffusive length scales (cf. Figures~3) has almost
no influence on the particle spectral index at the shock. Therefore
physical processes and/or conditions behind that distance are not
expected to modify the particle distribution at the shock.

The realistic conditions near astrophysical shocks are expected to
involve the diffusion coefficients depending on particle momentum. The
same will hold for the particle escape probability, defined by the
boundary distance in our model. Therefore, the present results can be
used only to a rather general evaluation of the acceleration conditions
and compatibility of the observed (possibly non-power-law) spectra and
time scales with the shock dynamics. A detailed modelling requires
numerical methods and not-frequently available information about the
local physical conditions and the boundary conditions.
 
Another fact of interest for modelling the generation of highest energy
particles near the spectrum cut-off should be mentioned in this place.
The first such particles to appear at the shock are those which have not
spent too much time diffusing far from the shock, in analogy to the case
with escape boundaries. Therefore, at a given energy, the time for these
particles to appear can be substantially shorter than the respective
acceleration time scale given in Equation~(1.2) or (2.24), the one valid
for the steady state particle spectrum. This fact is visible in analytic
time-dependent solutions (e.g. Drury 1991; our solution in Appendix A).
 
\thanks{We are grateful to the anonymous referee for valuable remarks.
This work was partly done during the visit of MO to Max-Planck- Institut
f\"{u}r Radioastronomie in Bonn. He is grateful to Prof. R. Wielebinski
and other colleagues from the Institute for their hospitality and
valuable discussions. The work of MO was supported from the KBN grant PB
1117/P3/94/06. RS acknowledges partial support by the Deutsche
Forschungsgemeinschaft (Schl 201/8-2). }

\section*{Appendix A: Inversion of Equation~(2.17) for $L_1=L_2=\infty $}

In the case of no escape boundaries $L_1=L_2=\infty $ we obtain from 
Equations~(2.13-16)

$$g_1(s,x,p)=G_1(s)G_2(s) \quad,\;\;\;\; g_2(s,x,p)=G_3(s)G_4(s) \quad
, \eqno (A1)$$

\noindent
with 

$$G_1(s)={3Q_0\over U_2(R-1)}\exp\left({2x\over \tau_1U_1}\right)\;
\left({p\over p_0}\right)^{-3/2}$$
$$\qquad H(p-p_0){\exp (A_1\sqrt{1+\tau_1s})\over s} \qquad , \eqno
(A2)$$

$$G_2(s)=\exp(-A_2\sqrt{1+\tau _2s})\qquad , \eqno (A3)$$
 
$$G_3(s)={3Q_0\over U_2(R-1)}\exp\left({2x\over \tau_2U_2}\right)\;
\left({p\over p_0}\right)^{-3/2}$$
$$H(p-p_0){\exp (A_3\sqrt{1+\tau_2s})\over s} \qquad , \eqno (A4)$$

$$G_4(s)=\exp(-A_4\sqrt{1+\tau _1s}) \qquad , \eqno (A5)$$

\noindent
where 

$$A_1={3R\over {2(R-1)}}\ln {p\over p_0}-\; {2x\over \tau_1U_1} \qquad ,
\eqno (A6)$$ 

$$A_2={3\over {2(R-1)}}\ln {p\over p_0} \qquad , \eqno (A7)$$

$$A_3={3\over {2(R-1)}}\ln {p\over p_0}+\; {2x\over \tau_2U_2} \qquad ,
\eqno (A8)$$

$$A_4={3R\over {2(R-1)}}\ln {p\over p_0} \qquad .  \eqno (A7)$$

\noindent
The upstream and downstream distribution functions then are 

$$f_1(t,x,p)=\int_0^tduF_2(u)F_1(t-u) \qquad  \eqno (A9a)$$

\noindent
and 
 
$$f_2(t,x,p)=\int_0^tduF_4(u)F_3(t-u) \qquad , \eqno (A9b)$$

\noindent
with the respective Laplace inverse transforms 

$$F_1(t)=L^{-1}\Bigl(G_1\Bigr)=
{3Q_0\over 2U_2(R-1)}\left({p\over p_0}\right)^{3\over 2(R-1)}
H(p-p_0)$$ 
$$\left\{
\hbox{erfc}\left(\left[{3R\over 4(R-1)}\ln (p/p_0)-\; {x\over
\tau_1U_1}\right]\sqrt{\tau_1/t}
+\sqrt{t/\tau_1}\right)\right.$$
$$+ \left({p\over p_0}\right)^{-{3R\over R-1}}
\exp\left({x\over \tau_1U_1}\right)
\hbox{erfc}\left(\left[{3R\over 4(R-1)}\ln (p/p_0)-
{x\over \tau_1U_1}\right]\sqrt{\tau_1/t} \right. $$
$$ \left. \left.
-\sqrt{t/\tau_1}\right)\right\} \qquad ,  \eqno (A10a)$$

$$F_2(t)=L^{-1}\Bigl(G_2\Bigr)=$$ 
$$ 
{{3\sqrt{\tau _2}\ln (p/p_0)}\over {4\sqrt{\pi }(R-1)t^{3/2}}} 
\exp\left(-{t\over \tau _2}-{{9\tau _2(\ln (p/p_0))^2}\over
{16(R-1)^2t}}\right) \quad ,  \eqno (A10b)$$
 
$$F_3(t)=L^{-1}\Bigl(G_3\Bigr)= 
{3Q_0\over 2U_2(R-1)}\left({p\over p_0}\right)^{3(2-R)\over 2(R-1)}
H(p-p_0)$$ 
$$\left\{
\left({p\over p_0}\right)^{-{3\over R-1}}
\hbox{erfc}\left(\left[{3\over 4(R-1)}\ln (p/p_0)+\; {x\over
\tau_2U_2}\right] \sqrt{\tau_2/t}
-\sqrt{t/\tau_2} \right)\right.$$
$$ \left. +
\exp\left({x\over \tau_2U_2}\right)
\hbox{erfc}\left(\left[{3\over 4(R-1)}\ln (p/p_0)+\; {x\over
\tau_2U_2}\right] \sqrt{\tau_2/t}
+\sqrt{t/\tau_2}\right)\right\} \, , \eqno (A10c)$$

$$F_4(t)=L^{-1}\Bigl(G_4\Bigr)=$$ 
$$ 
{{3\sqrt{\tau _1}R\ln (p/p_0)}\over {4\sqrt{\pi }(R-1)t^{3/2}}} 
\exp\left(-{t\over \tau _1}-{{9\tau _1R^2(\ln (p/p_0))^2}\over
{16(R-1)^2t}}\right) \quad .  \eqno (A10d)$$

\noindent
In deriving the four Laplace inversions (A10) we use the tables of
Oberhettinger and Badii (1973). Equations~(A9) and (A10) represent the
full solutions.
 
At the shock ($x=0$) we obtain
\vfill \break

$$f_0(p,t)=f_{1,2}(x=0,p,t)= 
{{9Q_0\sqrt{\tau _2}\ln (p/p_0)}\over {8\sqrt{\pi }U_2(R-1)^2}} 
\left({p\over p_0}\right)^{3\over 2(R-1)}
$$ 
$$H(p-p_0)\int_0^tdu u^{-3/2} 
\exp\left(-{u\over \tau _2}-{{9\tau _2(\ln (p/p_0))^2}\over
{16(R-1)^2u}}\right)$$
$$\left\{
\hbox{erfc}\left(\left[{3\over 4(R-1)}\ln (p/p_0)\right]\sqrt{\tau_1/(t-u)}
+\sqrt{(t-u)/\tau_1}\right)\right. +$$
$$+\left. \left({p\over p_0}\right)^{-{3R\over R-1}}
\hbox{erfc}\left(\left[{3R\over 4(R-1)}\ln
(p/p_0)\right]\sqrt{\tau_1/(t-u)}
-\sqrt{(t-u)/\tau_1}\right)\right\} \, .  \eqno (A11)$$

\noindent
After infinite time Equation~(A11) approaches the steady-state solution

$$f_0(p,t\to  \infty )\to {{9Q_0\sqrt{\tau _2} \ln(p/p_0)} \over
{4\sqrt{\pi }U_2(R-1)^2}} \left({p\over p_0}\right)^{{3\over 2(R-1)}
- {3R \over R-1}} $$
$$H(p-p_0)\int_0^\infty du u^{-3/2} \exp\left(-{u\over \tau _2}-{{9\tau
_2(\ln (p/p_0))^2}\over {16(R-1)^2u}}\right)$$
$$={{3Q_0H(p-p_0)}\over U_2(R-1)} \left({p\over p_0}\right)^{{-3R\over
R-1}} \qquad ,  \eqno (A12)$$

\noindent
where we used the integral

$$\int_0^{\infty }dx x^{-3/2}\exp \left( -x-{z^2\over 4x}\right)
=2^{3/2}K_{1/2}(z)z^{-1/2}= 2\sqrt{\pi }e^{-z}/z  \,\, . \eqno (A13)$$

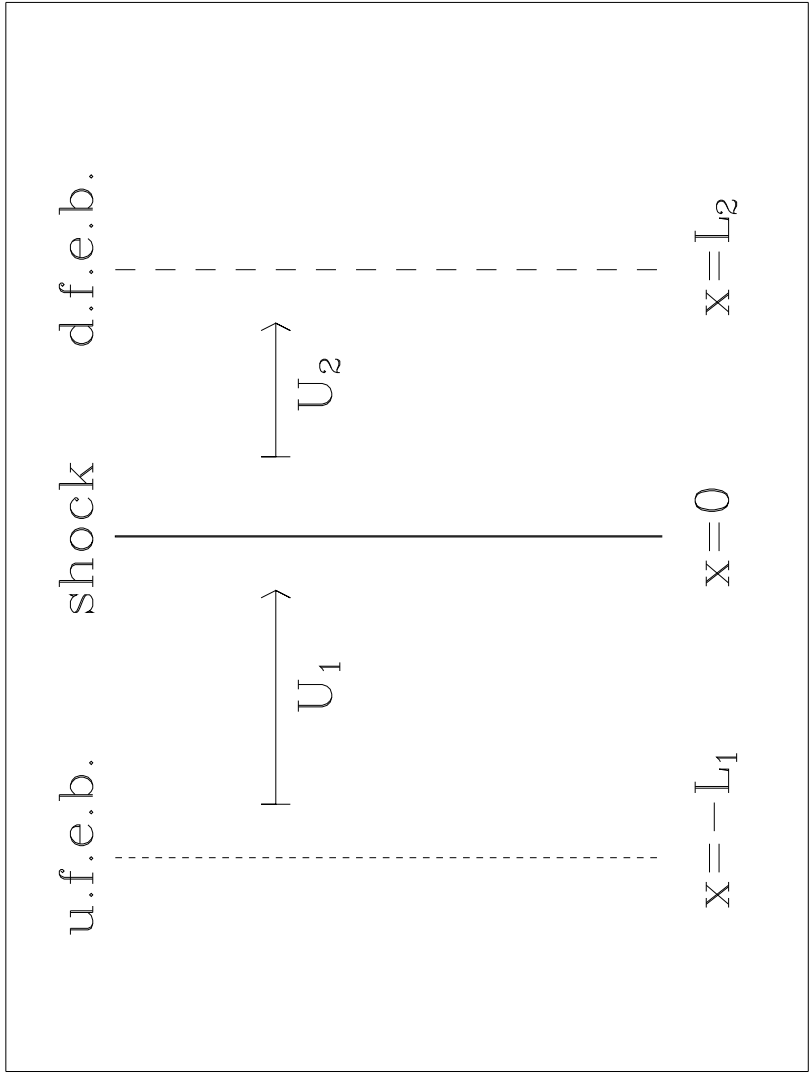
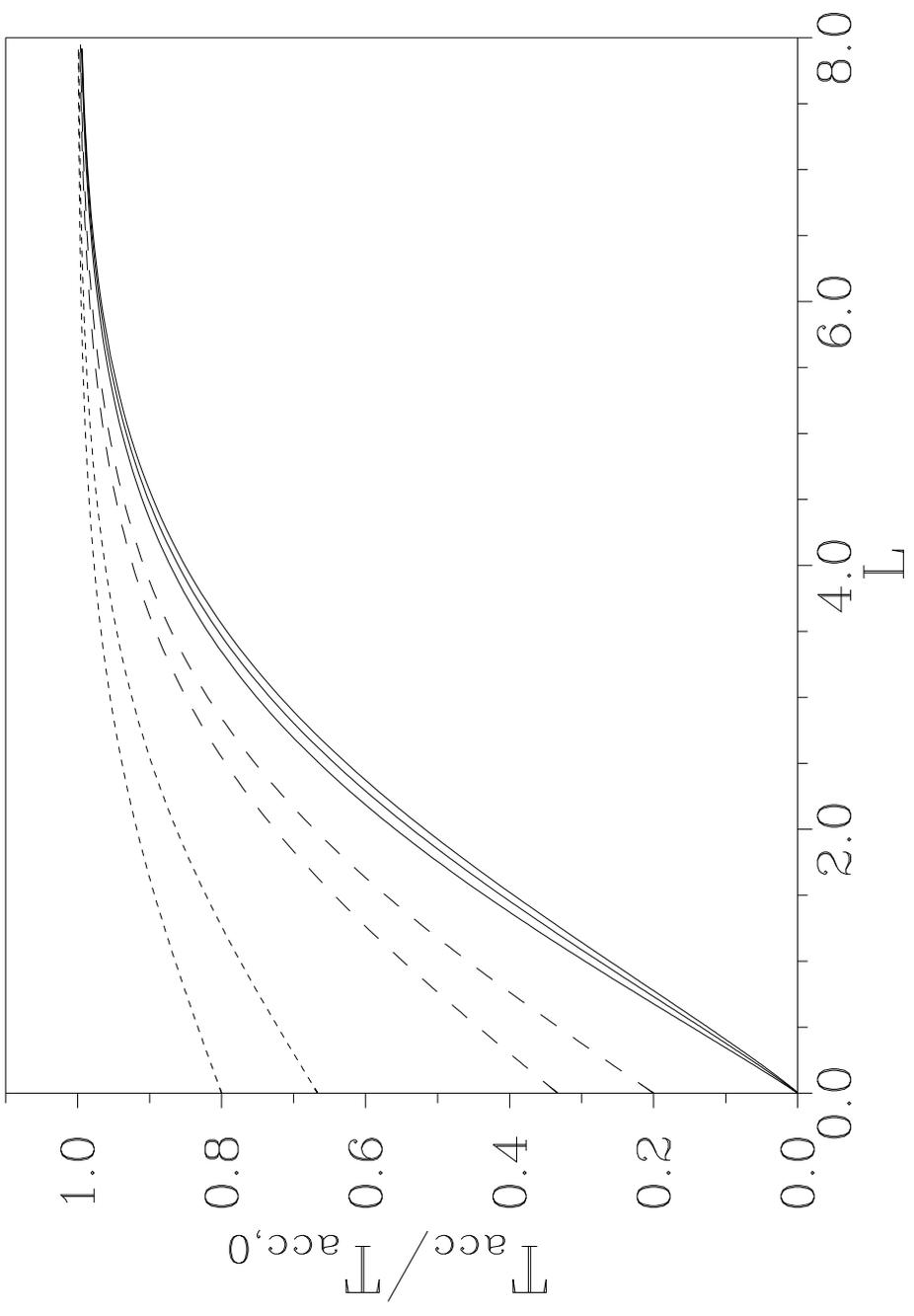
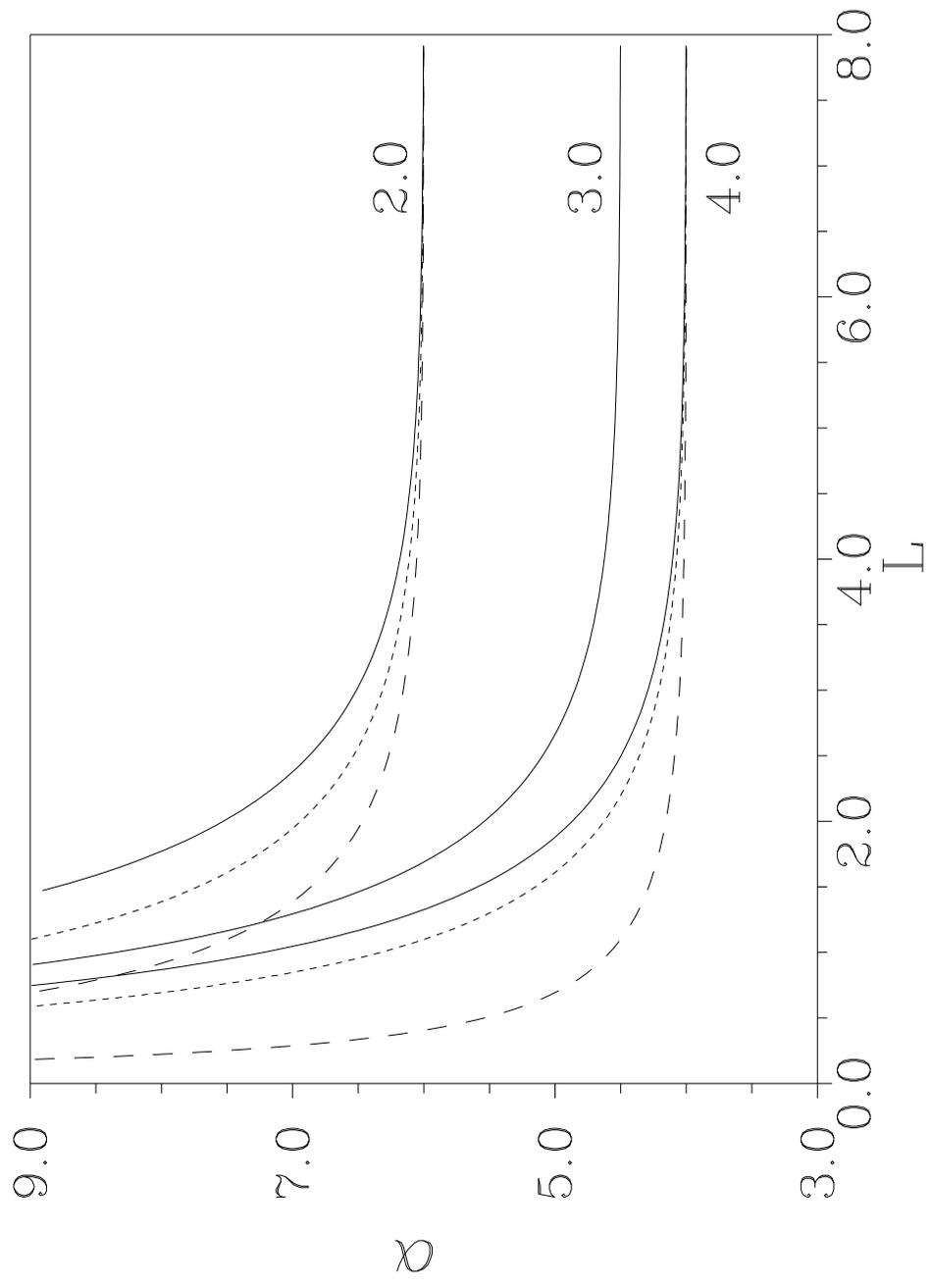
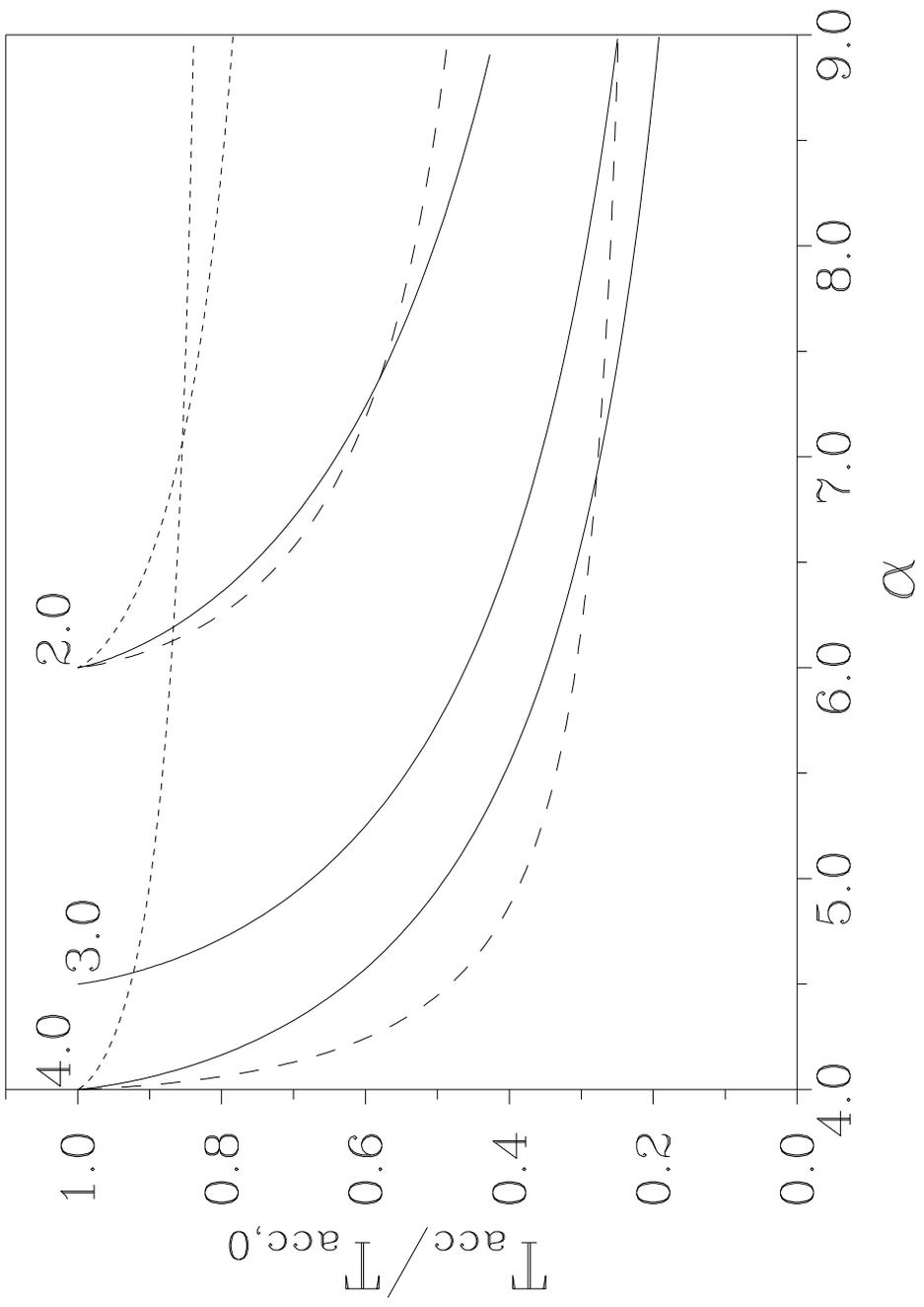
\end{document}